\documentclass[journal=jacsat,manuscript=article]{achemso}

\usepackage[version=3]{mhchem} 
\usepackage[T1]{fontenc}       
\usepackage{braket}

\author{Jamison Sloan}
\author{Nicholas Rivera}
\author{Marin Solja\v{c}i\'{c}}
\author{Ido Kaminer}
\affiliation{MIT Department of Physics, Cambridge, MA}
\alsoaffiliation{Technion University, Haifa, Israel}
\email{jamison@mit.edu}

\title{Tunable UV-Emitters through Graphene Plasmonics}

\abbreviations{IR,NMR,UV}

\begin{document}

\begin{tocentry}

\includegraphics{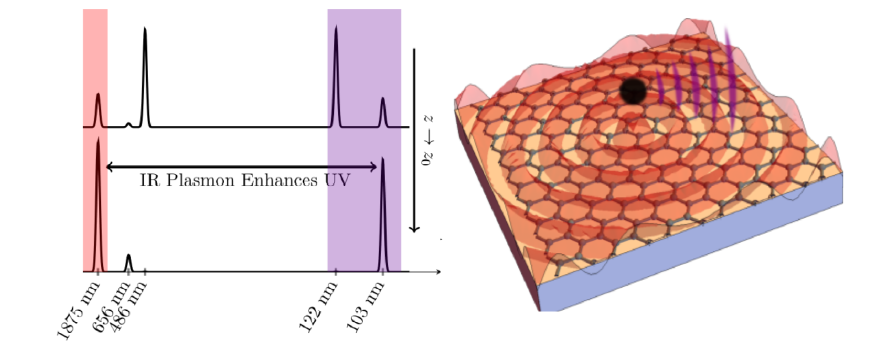}

\end{tocentry}

\begin{abstract}
Control over the spontaneous emission of light through tailored optical environments remains a fundamental paradigm in nanophotonics. The use of highly-confined plasmons in materials such as graphene provides a promising platform to enhance transition rates in the IR-THz by many orders of magnitude. However, such enhancements involve near-field plasmon modes or other kinds of near-field coupling like quenching, and it is challenging to use these highly confined modes to harness light in the far-field due to the difficulty of plasmonic outcoupling. Here, we propose that through the use of radiative cascade chains in multi-level emitters, IR plasmons can be used to enhance far field spectra in the visible and UV range, even at energies greater than 10 eV. Combining Purcell-enhancement engineering, graphene plasmonics, and radiative cascade can result in a new type of UV emitter whose properties can be tuned by electrically doping graphene. Varying the distance between the emitter and the graphene surface can change the strength of the far-field emission lines by two orders of magnitude. We also find that the dependence of the far-field emission on the Fermi energy is potentially extremely sharp at the onset of interband transitions, allowing the Fermi energy to effectively serve as a ``switch'' for turning on and off certain plasmonic and far-field emissions.
\end{abstract}

\section{Introduction}

One of the most fundamental results of quantum electrodynamics is that the spontaneous emission rate of an excited electron is not a fixed quantity; rather, it is highly dependent on the optical modes of the surroundings \cite{dirac1927quantum, tame2013quantum, pelton2015modified,cohen1992atom,scheel2008macroscopic,hoang2015ultrafast}. This result, known by most as the Purcell effect \cite{purcell1946spontaneous}, is the basis for the active field of spontaneous emission engineering, which has become paradigmatic in nanophotonics and plasmonics. 

One system that has emerged as a promising platform for studying strong plasmon induced light-matter interactions is graphene, which features low-loss, extremely sub-wavelength infrared surface plasmons with a dynamically tunable dispersion relation \cite{nagao2001dispersion, diaconescu2007low, liu2008plasmon, rugeramigabo2008experimental, jablan2009plasmonics, fei2012gate,woessner2014highly,zhang2014graphene,yan2013damping,gaudreau2013universal,archambault2010quantum,PhysRevLett.105.016801}. The combination of these features makes graphene prime for a wide array of applications such as tunable perfect absorbers, x-ray sources with tunable output frequency, tunable phase shifters with $2\pi$ phase control, tunable Casimir forces for mechanical sensing \cite{muschik2014harnessing}, tunable light sources via the plasmonic Cerenkov effect \cite{kaminer2016efficient}, and electrical control over atomic selection rules by taking advantage of access to conventionally forbidden transitions \cite{rivera2016shrinking}. Additionally, because graphene plasmons can be confined to volumes $10^8$ times smaller than that of a diffraction-limited photon \cite{nikitin2016real}, an infrared emitter in the vicinity of graphene can experience extreme enhancement of spontaneous decay through both allowed and forbidden channels via the Purcell effect \cite{koppens2011graphene,rivera2016shrinking}.

The Purcell effect and its consequences are almost universally studied in the framework of two-level systems. Surprisingly, the indirect effect of Purcell engineering on radiative cascade dynamics has rarely been exploited. Realistic emitters of course have many levels, and it is therefore possible to influence the spontaneous emission spectrum of an atom far beyond what a simple two-level analysis reveals, as we extensively make use of in this work. In particular, we find that by interfacing the (electrically and chemically) tunable IR Purcell spectrum of 2D plasmonic materials like graphene with multilevel emitters, it may be possible to design an electrically tunable UV frequency emitter, even when no plasmons exist at those frequencies. Remarkably, we find that it is possible to use the Purcell effect at IR frequencies to enhance far-field emission of 100 nm wavelength light by nearly two orders of magnitude. 

\section{Methods}

In our work, we consider the spectrum produced by emitters near tunable plasmonic environments such as graphene. For calculational concreteness, we take a hydrogenic emitter, whose spectrum shares many features with more generic atoms. The hydrogen atom has a set of electronic states indexed by quantum numbers $\ket{n,l,m_l,m_s}$, where $n$ is the principal quantum number, $l$ is the orbital quantum number, $m_l$ gives the orbital angular momentum, and $m_s$ is the spin of the electron. The effect of spin-flip transitions is negligible for our purposes, and thus we will not keep track of the electron spin quantum number $m_s$. Additionally, we will not consider fine structure splitting, so the energies of states are indexed solely by the principle number $n$. Electrons have the ability to transition between states through emission and absorption of light. Strictly speaking, the electromagnetic fluctuations of the system are mixtures of far-field photons and near-field plasmons, but in practice, every excitation can be well-defined as either a photon or a plasmon to a good approximation. Here, excitations at IR frequencies will be considered purely plasmonic in nature, while excitations in the visible and UV will be treated as far-field photons.

We consider a multi-state system with a pump from the ground state $\ket{\text{g}}$ of Hydrogen to some excited state $\ket{\text{e}}$. Both optical and electrical pumping could be considered. Once excited, the electron can then radiatively cascade back down to the ground state, emitting photons of one or more frequencies in the process. The dynamics of the system are governed by the rate equation 
\begin{equation}
  \frac{d\mathbf{N}}{dt} = A\mathbf{N},
  \label{eq:rate_equation}
\end{equation}
where $\mathbf{N}$ is a vector of length $n$ containing the occupation numbers of the $n$ electronic states, and $A$ is a rate matrix with entries $A_{ij} = (1-\delta_{ij})\Gamma_{ij} - \delta_{ij}\sum_{k=1}^n \Gamma_{kj}$, where $\Gamma_{ij}$ is the rate of transition between states $i$ and $j$. The rate $\Gamma_{ge}$ describes the pumping of photons from the ground state to the starting excited state. The rates of all upward transitions with the exception of the pump $\Gamma_{\text{ge}}$ are assumed to be zero. After sufficient time, a steady state equilibrium is established between the pump and the cascading photons, i.e. $\frac{d\mathbf{N}_s}{dt} = A\mathbf{N}_s = 0$, where $\mathbf{N}_s$ contains the steady state populations of the emitter levels.

The total rate of production of a particular frequency photon is obtained by summing over all channels of the same frequency of emission to account for potential degeneracies. That is,
\begin{equation}
  \frac{dp_\omega}{dt} = \sum_{\omega_{ij} = \omega}\Gamma_{ij}N_i,
\end{equation}
where $\omega_{ij} = \omega_j - \omega_i$. Then, the observed power output of a frequency $\omega$ is given as
\begin{equation}
  P(\omega) = \hbar\omega\frac{dp_\omega}{dt}.
  \label{eq:power_spectrum}
\end{equation}
The observed differential spectrum $dP(\omega)/d\omega$ will be subject to broadening effects, such as doppler broadening and inhomogeneous broadening which we do not consider here. This model is easily extended by adding non-radiative loss channels directly into the rate equations. Also note that we assume electrons in the 2s state return to the ground state without the emission of a photon since this effect is negligible at first order \cite{shapiro1959metastability}.

The spontaneous emission rate from state $\ket{i}$ to $\ket{j}$ near the surface of graphene is given as $\Gamma_{ij} = \Gamma_0F_p(\omega)$, where $\Gamma_0$ is the rate of transition in the vaccum, and $F_p(\omega)$ is the Purcell factor. The Purcell factor for p-polarized modes is given as\cite{koppens2011graphene}
\begin{equation}
  F_p(\omega) = 1 + f \frac{3c^3}{2\omega^3}\int q^2 \exp(-2qz_0)\text{Im}\left[\frac{1}{1 - q\sigma(q,\omega)/2i\omega\epsilon_0}\right]\,dq,
  \label{eq:purcell_factor}
\end{equation}
where $\omega$ is the plasmon frequency, $q$ is the plasmon in-plane wavevector, $z_0$ is the distance between the graphene and the emitter, $\sigma(q,\omega)$ is the conductivity of graphene, and $f = 1\,(1/2)$ for dipoles perpendicular (parallel) to the graphene plane. In Figure 1, we see such an emitter near the surface of graphene that is able to radiate into plasmonic surface modes as well as into the far-field. We also see the energy levels of hydrogen with the possible decay pathways from the 4d state shown as given by the dipole selection rules. By calculating the rates, we compute the power spectrum, which we claim can undergo a drastic shift as the emitter is brought into sufficient proximity of the graphene surface.

In our study, we consider two models of conductivity: the Drude model, $\sigma_D(\omega) = i(e^2 E_F/\pi\hbar^2)/(\omega + i\tau^{-1})$, and the local interband conductivity $\sigma(\omega) = \sigma_D(\omega) + \sigma_I(\omega)$, where the contribution from interband effects is $ \sigma_I(\omega) = \frac{e^2}{4\hbar}\left(\theta(\hbar\omega - 2E_F) - \frac{i}{\pi}\ln\left|\frac{2E_F + \hbar\omega}{2E_F - \hbar\omega}\right|\right)$\cite{jablan2013plasmons}. In the above, $E_F$ is the Fermi energy of the graphene substrate, which is directly related to the electron carrier density, and $\tau$ is the empirical relaxation time corresponding to losses that can generally be a function of frequency and vary with the Fermi energy \cite{jablan2013plasmons}. In this work we neglect the dependence of $\tau$ on $E_F$ but it can be accounted for using results of density functional theory analysis.\cite{papadakis2017ultra} While the local model is more precise and has been demonstrated to well-describe flourescence quenching experiments in graphene \cite{tielrooij2015electrical}, we also consider the Drude model to connect to other 2D metals and also other Drude metals featuring high local density of states. We comment on the neglect of nonlocality later in the text.properly

\section{Results and Discussion}

Using the rate equation formalism of the previous section, we now explore how proximity to graphene can enhance atomic spectra through radiative cascade. We consider a hydrogen atom pumped from the $1s$ state to a $4d$ state. In order to understand the dynamics of enhancement at work, we consider the Purcell factors which enhance each transition frequency at various distances from emitter to the graphene surface. In Figure 2(a) we show the Purcell factor given in Eq.~\ref{eq:purcell_factor} for a dipole at frequecy $\omega$ at a distance $z_0$ from a graphene surface doped to $E_F = 1.0$ eV using the Drude model of conductivity. Losses are taken into an account with Drude relaxation time of $\tau = 10^{-13}$ s. At low frequency and low $z_0$, loss induced quenching causes the Purcell factor to exhibit divergent behavior. At mid frequency range 0.05 - 0.8 eV, the Purcell enhancement comes primarily from plasmonic emission, and can easily reach $10^6$ at experiementally realizable $z_0$ such as 5 nm. At sufficiently high frequencies ($\geq 0.8$ eV), proper plasmonic modes cease to exist, and the relevant electromagnetic fluctuations instead can couple to particle-hole excitation. In this region, the supported modes are simply the far-field free space modes. Panel (2c) shows the Purcell factor using the full local RPA model of graphene conductivity which accounts for interband transitions. The main difference between the Drude and full local RPA model can be seen in the $10^5 - 10^6$ Hz (0.66 - 6.6 eV) frequency range. The RPA model exhibits a sharp dip in the Purcell factor at a critical frequency characterized by the condition $2E_F = \hbar\omega$, corresponding to the divergence of the imaginary part of the conductivity.

Panels (b) and (d) show the rate of photon production for a hydrogenic dipole emitter approaching the surface of graphene, calculated for the conductivity models described in (a) and (c) respectively. When the emitter is placed within nanometers of the graphene sheet, infrared plasmon emission is enhanced by the Purcell effect. In fact, the direct plasmonic enhancement of IR transitions is much greater than that of vis-UV transitions, leading to dominance of IR decay pathways in the presence of graphene. This alteration of the decay pathways can induce substantial modification of the far-field spectrum. As an example, at distances closer than 20 nm the $3p \to 1s$ (103 nm) UV transition dominates the $2p \to 1s$ transition (121 nm) which is normally prominent in free space. Note that this dominance comes not from direct enhancement of the local density of states at the $3p\to 1s$ transition frequency, but rather the enhancement of the IR transition $4d\to 3p$ at 1875 nm associated with plasmon emission. The enhanced IR transition populates $3p$ with electrons which then prefer to decay into the $1s$ state. We note that this principle is easily extended to other emitters with much higher frequency transitions (such as helium with a 30+ eV transition and perhaps even EUV transitions). It would also be of interest to extend this idea to higher energy core-shell transitions in heavier atoms.

In Figure 3, we see the calculated spectral power output of the emitter at four different distances $z$. At a distance $z=100$ nm shown in panel (d), the emitter is too far from the surface of graphene to couple to plasmonic modes, so the output spectrum is effectively that of an emitter in free space. As the emitter nears the surface, IR plasmons are excited at 1875 nm, re-routing the power output of the emitter into the 103 nm UV channel. The strengthening of the 103 nm line can be seen at distance $z=20$ nm shown in panel (c). At distances below 10 nm, more than 90\% of the power output is directed into the 103 nm channel. Both panels (a) and (b) show the vast majority of spectral output in the 103 nm line. Indirect coupling of IR and UV transition rates is a highly efficient method of UV enhancement, as an order of magnitude or more power is channeled into the far-field UV emission than the excitation of the supporting IR plasmon. 

We now exploit features of the full local conductivity model to work toward a spectrum that can be both drastically modified and delicately tuned by doping graphene. As we recall from Fig. 2(c), the Purcell factor calculated in the full local conductivity model exhibits a sharp dip near $2E_F = \hbar\omega$ as a result of the corresponding logarithmic singularity in $\sigma_I(2E_F/\hbar)$. By tuning the Fermi energy such that $2E_F/\hbar$ corresponds to a characteristic transition frequency $\omega_0$ of the emitter, the rates of other transition frequencies can be greatly enhanced relative to that of frequency $\omega_0$. What results is a dramatic relative slowing of the $\omega_0$ transition, which can cause reduced intensity not only in the emission line $\omega_0$, but also in emission lines corresponding to transitions enabled by radiative cascade after the $\omega_0$ transition. 

In Figure 4, we see the photon emission rates at four distances $z_0$ as a function of varying Fermi energy $E_F$. The two lowest energy transition in this system are at 1875 nm and 656 nm, corresponding to critical $E_F$ values of 0.34 eV and 0.95 eV respectively. Crossing these boundaries causes critical changes in the branching ratios of the system. For example, at a distance of $z_0=20$ nm shown in panel (d), a Fermi energy of 0.34 eV suppresses the $4d \to 3p$ (1875 nm) transition, whereas a Fermi energy of 0.32 eV allows the 1875 nm transition to dominate. Since electrons in the $3p$ state are far more likely to transition into $1s$ than $2s$, crossing this Fermi energy boundary acts not only as a ``switch'' for the 1875 line, but also one which amplifies the 103 nm UV line, and suppresses the visible line at 486 nm. As the emitter nears the surface at distances of $z_0 = 1$ nm or 5 nm, as shown in panels (a) and (b), the system dynamics change. Namely, with increasing proximity to the surface, the 1875 nm line remains dominant for most values of $E_F$ while the 103 nm line decreases in intensity. However, after crossing the critical threshold corresponding to suppression of the 656 nm line, the 103 nm line is once again enabled to match the intensity of the 1875 nm channel. As another example, consider that by changing the distance from $z=5$ nm to $z=10$ nm, one can change which channel the 656 nm line follows. At $z_0=5$ nm (b), the 656 nm line matches the rate of the 1875 nm line for $E_F$ below the 656 nm suppression threshold. In contrast, at $z_0=10$ nm (c), the 656 nm line instead follows the 103 nm line. We see that crossing critical doping boundaries allows significant modification of the spectral structure, while tuning between critical points allows smooth and controlled deformation.

Using the full nonlocal RPA conductivity model, one can estimate the effects of nonlocality on our calculations.\cite{koppens2011graphene} The effects of nonlocality are most significant at low Fermi energies. In these regimes, the Purcell factors near $\omega = 2E_F/\hbar$ can become larger than in the local approximation by around 2 orders of magnitude. The result is a less drastic but stil critical behavior at the expected points which should be observable. At $E_F > 0.5$ eV, the nonlocal corrections are comparatively much smaller. Additionally, note that nonlocal considerations should not significantly impact the spectrum as a function of distance $z$ variation so long as no important transitions lie in the conductivity divergent frequency range. In other words, strong indirect enhancement of far-field transitions should still be achievable, even in the presence of nonlocal effects.

From the analysis here, it is clear that the effect of the Fermi energy of graphene serves not only as a knob to alter plasmonic coupling, but also one which can modify coupling into far-field modes where there are no plasmons - all due to the radiative cascade effect which effectively correlates the emission of IR and UV frequencies. By varying both the Fermi energy and the distance to substrate $z$, a wide variety of spectral regimes can be accessed. Changing either of these paramaters has the ability to change the fundamental behavior of the emitter. As another example, consider that by changing the distance from $z=5$ nm to $z=10$ nm, one can change which channel the 656 nm line follows. At $z_0=5$ nm (b), the 656 nm line matches the rate of the 1875 nm line for $E_F$ below the 656 nm suppression threshold. In contrast, at $z_0=10$ nm (c), the 656 nm line instead follows the 103 nm line. 



Excitation of the emitter to higher energy level states can greatly increase the number of decay pathways available to an electron, as we demonstrate in Figure 5. As an example, consider an emitter excited to the state $5p$. At distances $z_0 = 20$ nm from the surface (c), the emission rate of 95 nm photons into the far-field can increase by a factor of 50 or more across a wide range of carrier densities. In particular, the critical threshold corresponding to the 4050 nm plasmon has the effect of exchanging the spectral dominance of the 95 nm and 103 nm lines. In this case, bringing the atom much closer than 10 nm from the surface changes decay dynamics in a way which actually suppresses this effect, which can be seen in panel (a). At a distance as close as 5 nm, plasmons of lower frequency than those which enable the fast UV transition become excited, competing with the desired high frequency transitions. Perhaps counterintuitively, the strongest enhancement of UV far-field emission occurs at moderate, rather than extremely close or extremely far distances of emitter to surface. At extremely close distances to the surface, plasmons of higher frequency become excited, working against the ability of very low frequency IR plasmons to focus electrons through a very specific decay channel. This emphasizes that strong high frequency enhancement requires not only the enhancement of IR plasmons which enable UV transitions, but also the relative dominance of this enabling IR transition to other competing mechanisms of decay.  

\section{Conclusions and Outlook}

In summary, we have demonstrated that by modulating the carrier density and proximity of an emitter to graphene, the far-field emission spectrum can be fundamentally altered by greatly enhancing the rates of coupling to the electromagnetic fluctuations of the graphene sheet. We note that while we have emphasized the enhancement of UV emission, our results make it evident that the entire spectrum can be drastically altered. This includes both the partial or nearly complete suppression of ordinarily present spectral lines, as well as the amplification of lines that are slow under free-space conditions. Since UV transitions generating far field photons may be indirectly enhanced through cascade, it should be possible to observe evidence of enhancement in an experimental setting, even without the ability to outcouple plasmons.

Polaritons other than surface plasmon-polaritons on graphene also provide a broad range of scenarios in which similar effects can be observed, such as phonon and exciton-polaritons. For example, surface phonon-polaritons on hexagonal boron nitride can provide similar Purcell enhancement-type effects. In fact, the Restrahlen band of hBN could allow for even more selective enhancement of first or higher order decay mechanisms, enabling yet another mechanism of control over which transitions are enhanced \cite{rivera2016near}.

An alternate possibility for controlling the far-field spectrum is by not only engineering the Purcell enhancement of allowed dipolar transitions, but also utilizing forbidden transitions \cite{zurita2002multipolar,zurita2002multipolar2,takase2013selection,yannopapas2015giant,rukhlenko2009spontaneous,jain2012near}. As shown in \cite{rivera2016shrinking}, it should be possible to use the highly confined plasmons in graphene to make highly forbidden emitter transitions occur at rates competitive with those of electric dipole transitions. Then by radiative cascade, it should be possible to extract a far-field UV signal from these forbidden transitions. Our analysis presented here thus serves as a crucial starting point for designing experiments to detect transitions whose observations have proved elusive since the discovery of spontaneous emission.

\section{Figures}

\begin{figure*}[h!]
\centering
\includegraphics[width=0.9\textwidth]{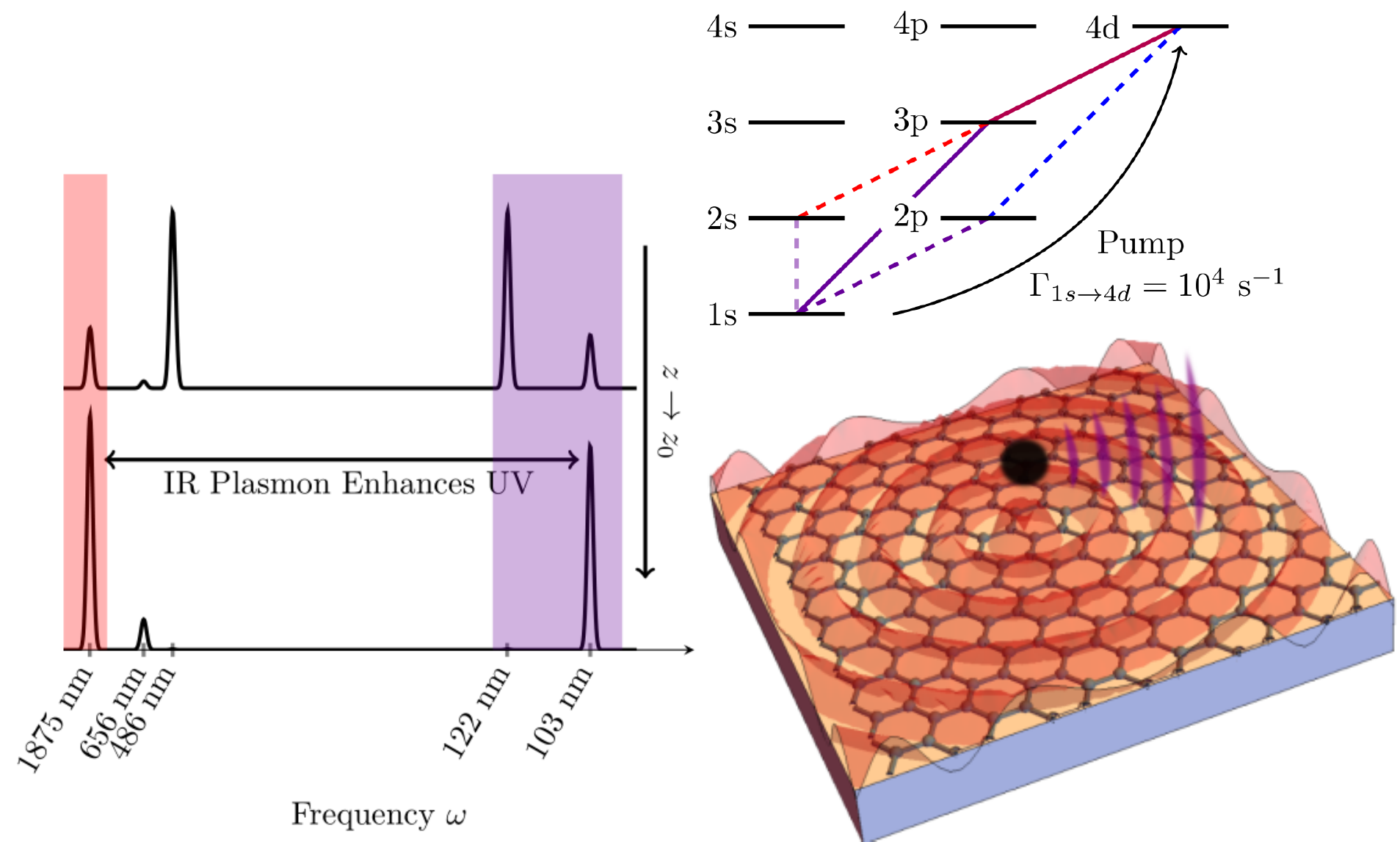}
\caption{Hydrogenic emitter near the surface of graphene with pumping laser. As the atomic emitter is brought closer to the graphene substrate, surface plasmons can form, enhancing the emission rate in the IR spectrum. Through radiative cascade, enhancement is also observed in the UV at 103 nm.}
\label{fig:schematic}
\end{figure*}

\newpage

\begin{figure*}[h!]
\centering
\includegraphics[scale=1.2]{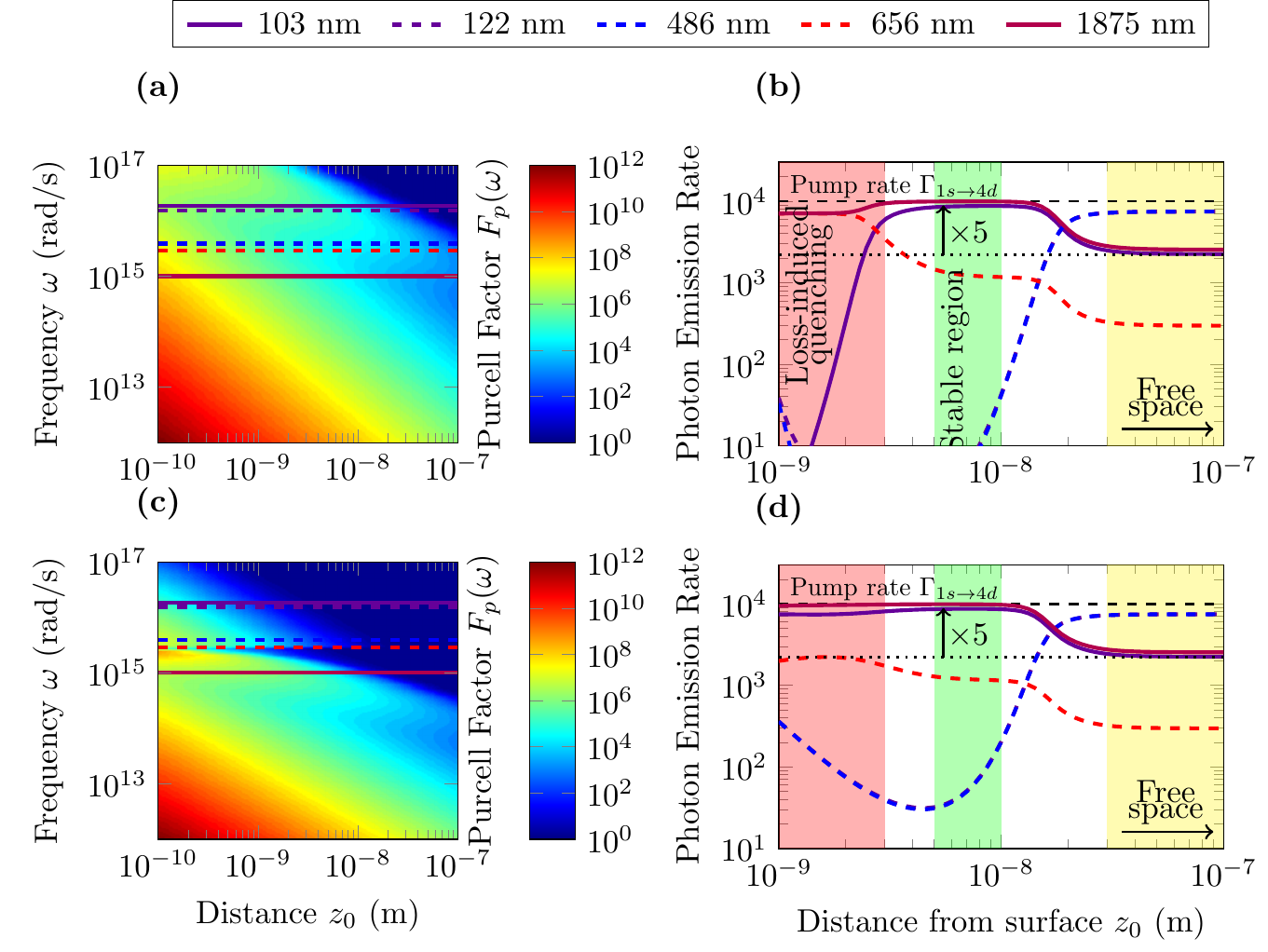}
\caption{Rate of photon emission of different wavelengths via radiative cascade for electrons pumped from 1s to 4d state of Hydrogen via laser at a rate of $\Gamma_{1s\to4d} = 10^4$ s$^{-1}$. Calculations performed at $E_F = 1.0$ eV and losses of $\tau = 10^{-13}$ s. Plots a) and c) show the Purcell factor as a function of distance and frequency in the a) Drude model regime and c) full interband conductivity model regime. Plots b) and d) show the rate of photon emission corresponding to these regimes. The colored lines in the Purcell factor plots correspond to the frequencies of emission shown in the probability plots.}
\end{figure*}

\newpage

\begin{figure*}[h!]
\centering
\includegraphics[scale=1.1]{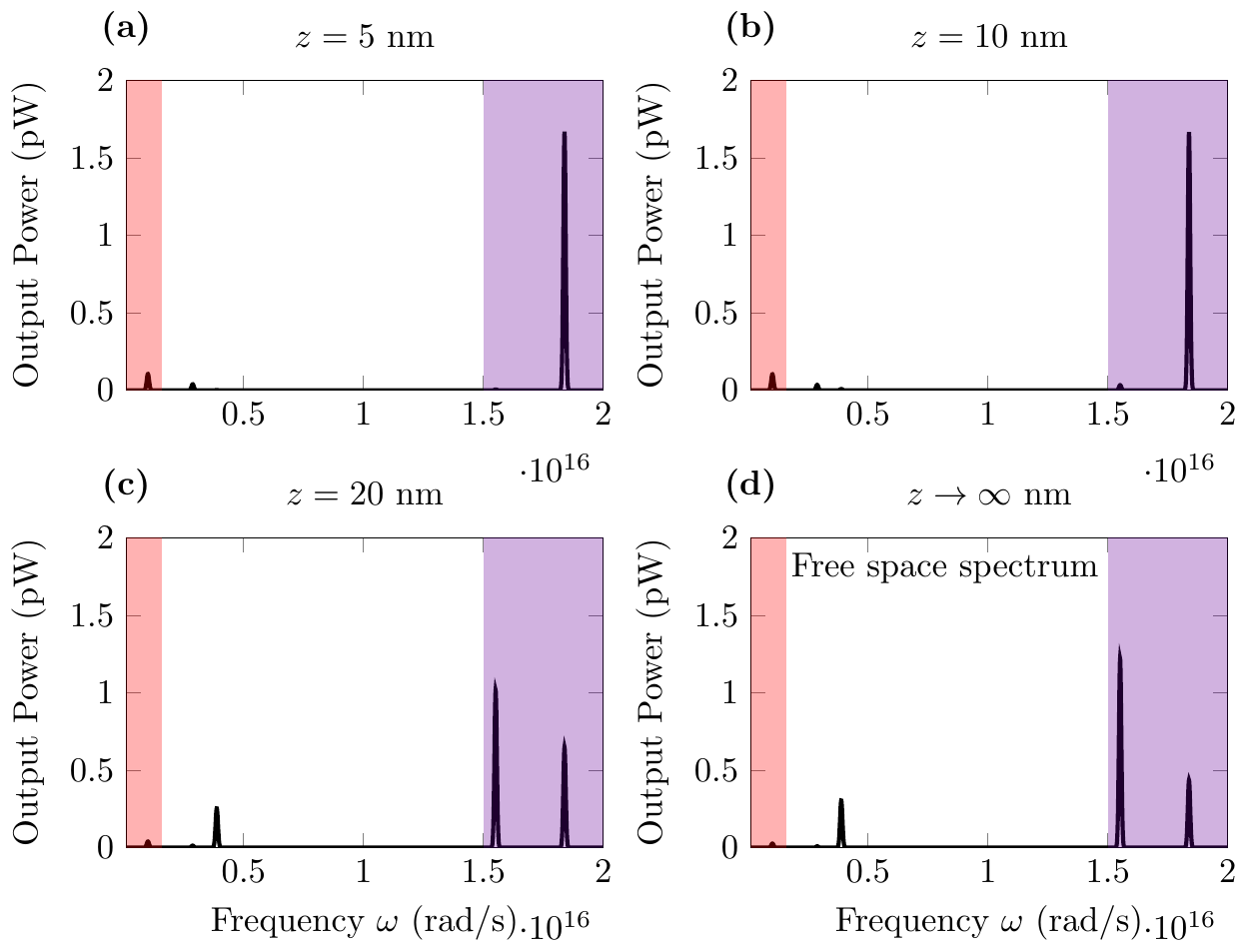}
\caption{Calculated power spectrum of Hydrogenic emitter at four distances $z$ from the surface of graphene. Calculations were done assuming a pump rate $\Gamma_{1s\to 4d} = 10^4$ s$^{-1}$, full local conductivity model with losses characterized by $\tau = 10^{-13}$s, and graphene surface doped to $E_F = 1.0$ eV. As the emitter approaches the surface, the 1875 nm transition in the IR is enhanced, amplifying the power output of the 103 nm UV line, while suppressing the strength of other spectral lines.}
\end{figure*}

\newpage

\begin{figure*}[h!]
\centering
\includegraphics[scale=1.1]{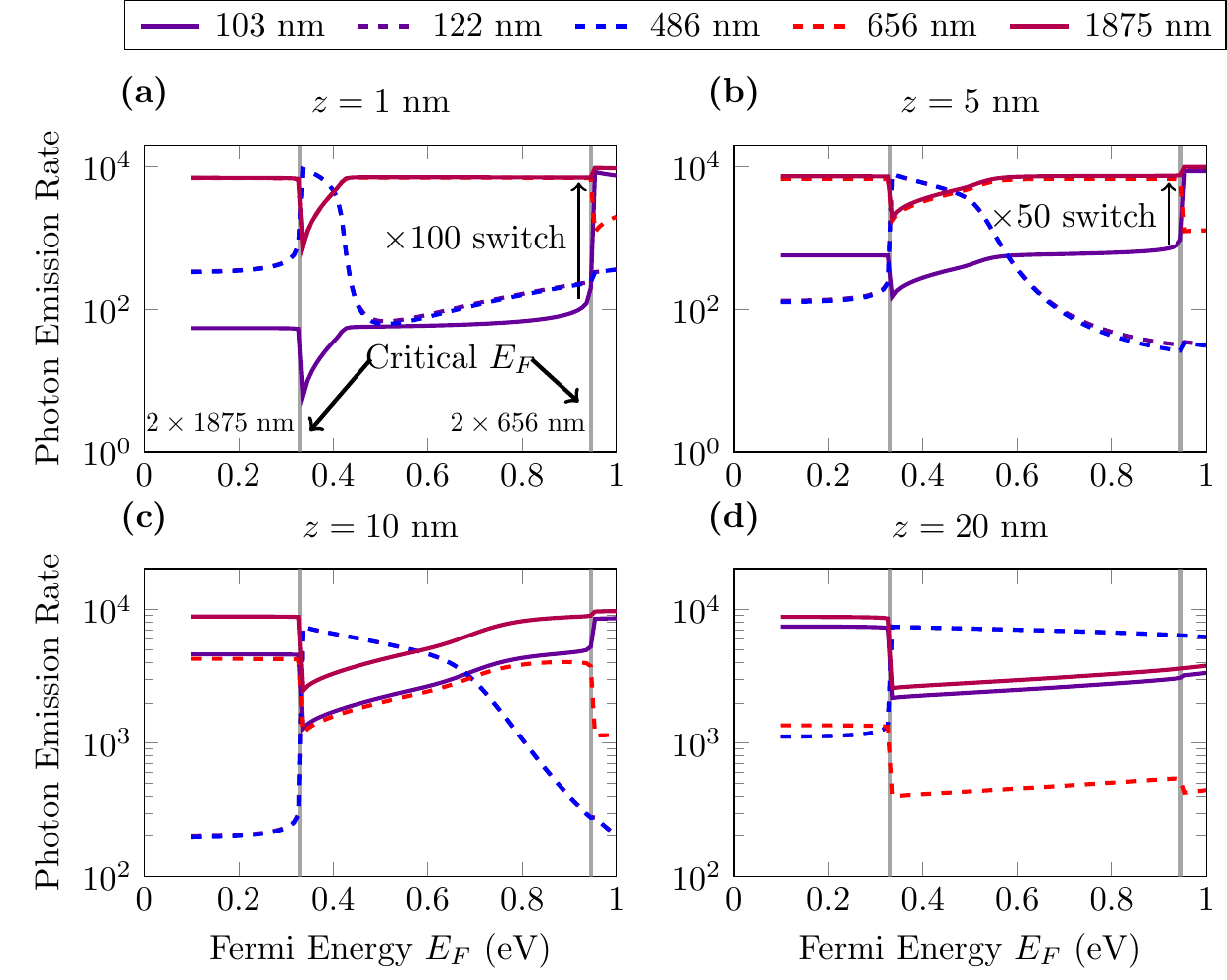}
\caption{Rate of emission of different wavelength photons via radiative cascade with 1s photons pumped to the 4d state of Hydrogen via laser ($\Gamma_{1s\to4d} = 10^4$ s$^{-1}$), with varying Fermi Energy $E_F$, and with graphene at distances $z=1,5,10,20$ nm from the emitter. Calculations are done using the full intraband conductivity model with losses characterized by $\tau = 1\times 10^{-13}$s. Divergence of the imaginary part of the full conductivity at $E_F = \hbar\omega/2$ results in highly critical behavior in the spectrum around $E_F$ which correspond by this relation to $\omega$ of transitions in the electronic spectrum. In particular, $2\times \lambda_{1875}$ nm corresponds to critical behavior for 1875nm, and $2 \times 656$ nm corresponds to critical behavior for 656nm. As the distance $z$ decreases, crossing critical points causes larger changes in the spectrum. Also note that at low $E_F$, emission rates stabalize.}
\end{figure*}

\newpage

\begin{figure*}[h!]
\centering
\includegraphics[scale=1.1]{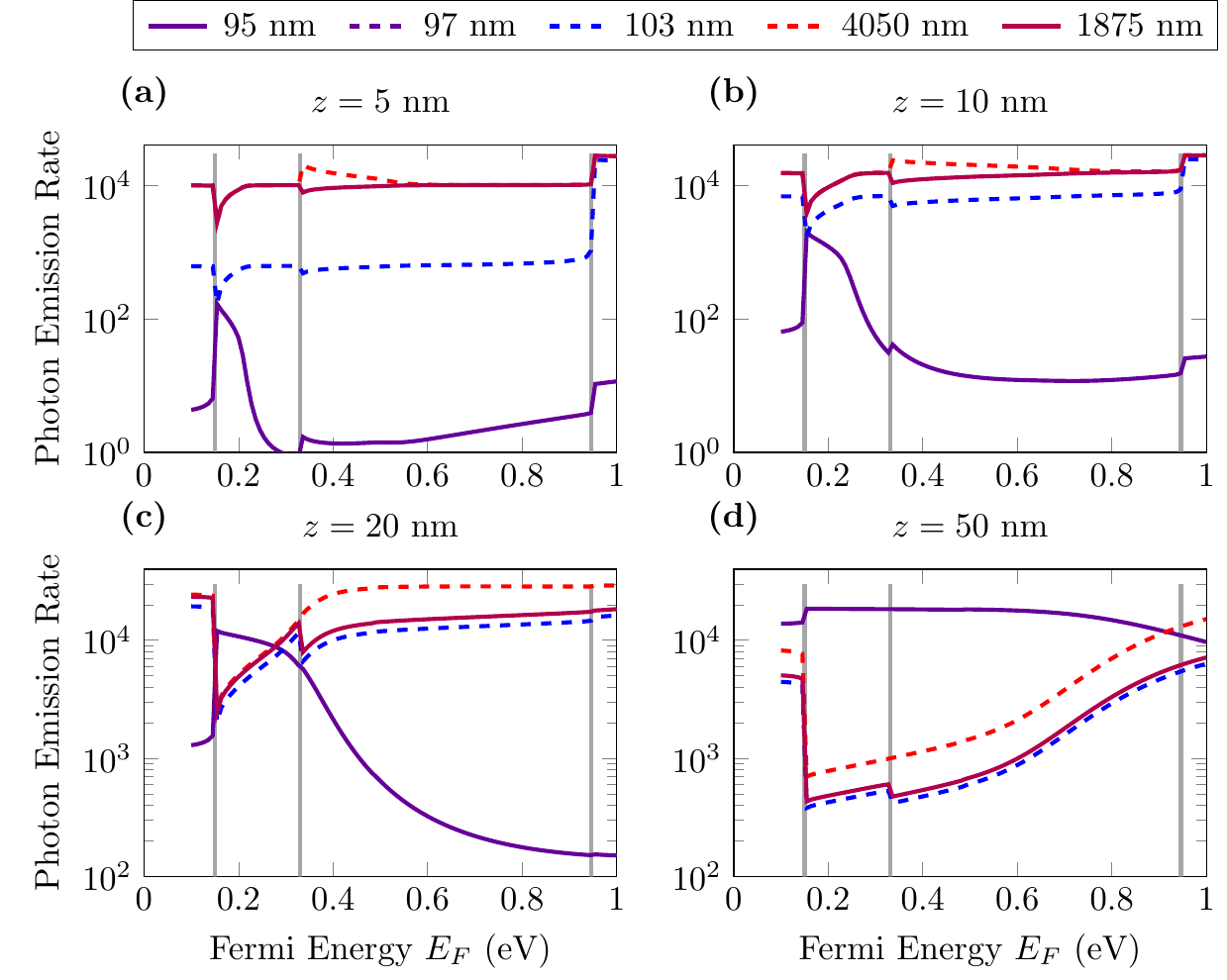}
\caption{Rate of emission of different wavelength photons via radiative cascae with $1s$ photons pumped to the $5p$ state of Hydrogen, with varying Fermi Energy $E_F$, and with graphene at distances $z=5, 10, 20, 50$ nm from the surface.}
\end{figure*}

\begin{acknowledgement}

Research supported as part of the Army Research Office through the
Institute for Soldier Nanotechnologies under contract
no. W911NF-13-D-0001 (photon management for developing
nuclear-TPV and fuel-TPV mm-scale-systems).
Also supported as part of the S3TEC, an Energy Frontier Research
Center funded by the US Department of Energy
under grant no. DE-SC0001299 (for fundamental photon
transport related to solar TPVs and solar-TEs).
shown in this document.

\end{acknowledgement}


\bibliography{thesisbib}

\end{document}